\documentstyle[11pt]{article}
\input amssym.def
\def\ra{\rangle}
\def\la{\langle}

\begin{document}
\baselineskip18pt
\thispagestyle{empty}
\begin{center}
{\LARGE\bf Separability of rank two quantum states on multiple
quantum spaces}
\end{center}
\vskip 2mm
\begin{center}
 {\normalsize Shao-Ming Fei$^{1,\  2}$, Xiu-Hong Gao$^1$, Xiao-Hong Wang$^1$,
 Zhi-Xi Wang$^1$, and Ke Wu$^1$}\\ {\small \sl $ ^1$ Department of Mathematics,
Capital  Normal University, Beijing, China.}\\ {\small \sl $ ^2$
Institute of Applied Mathematics, University of Bonn,  53115 Bonn
Germany}
\end{center}

\vskip 2mm
\parbox{14cm}
{\footnotesize{\bf Abstract}    Explicit sufficient and necessary
conditions for separability of $N$-dimensional rank two multiparty
quantum mixed states are presented. A nonseparability inequality
is also given, for the case where one of the eigenvectors
corresponding to nonzero eigenvalues of the density matrix is
maximally entangled.}

\vskip 4mm

Quantum entanglement is playing an essential role in quantum
information processing (cf. \cite{effects,geste,Tel}). The
separability of pure states for bipartite systems is quite well
understood (cf. \cite{peresbook}). For mixed states, some progress
has been achieved in understanding the separability and
entanglement problem for bipartite systems (cf. \cite{primer}),
e.g., the proper definition of separable and entangled states
formulated by Werner \cite{Werner}, the Peres \cite{Peres}
criterion that all separable states necessarily have a positive
partial transpose, which is further shown to be also a sufficient
condition for separability in $2\times2$ and $2\times 3$ systems
\cite{ho96,tran}.

The multiparty entangled states have been investigated recently
\cite{rew2000,four}. These states are also of importance in
quantum information processing, for instance, three party
entanglement of the GHZ type can allow for interesting
applications such as quantum secret sharing \cite{sharing}, and
many experimental groups have tried to generate such states
\cite{exper}. However less is known about how to characterize
multiparty entanglements completely. The maximum connectedness,
persistency \cite{persis} and Schmidt measure \cite{schmidt} have
been used in partly describing entanglements of pure multiparty
states. There exists no general criterion that allows one to
distinguish whether a general mixed state is separable or not. In
\cite{22n} the separability and entanglement in $2\times 2\times
N$ composite quantum systems have been studied.

In this paper, we study sufficient and necessary conditions for
separability of higher-dimensional quantum systems on $H\otimes
H\otimes\dots\otimes H$, which generalize the results in \cite{4}.
In particular, we consider density matrices with rank two. The
separability condition for these kind of mixed states in arbitrary
dimensions is explicitly given. In addition, we present a
nonseparability inequality valid in the case where one of the
eigenvectors corresponding to nonzero eigenvalues of a density
matrix is maximally entangled.

We first consider the case of three $N$-dimensional quantum
spaces. Let $H$ be an $N$-dimensional complex Hilbert space, with
$e_i$, $i=1, 2, \dots, N$, an orthonormal basis. A general pure
state on $H\otimes H\otimes H$ is of the form
\begin{equation}
\label{1} |\Psi\ra =\sum_{i,j,k=1}^N a_{ijk}e_i\otimes e_j\otimes
e_k, \hspace{0.2cm} a_{ijk}\in {\Bbb C}
\end{equation}
with the normalization $\sum _{i,j,k=1}^N a_{ijk}a_{ijk}^*=1$ (*
denoting complex conjugation).

Let $U$ denote a unitary transformation on the Hilbert space $H$,
such that
$$ Ue_i=\sum_{j=1}^N b_{ij}e_j, \hspace{0.2cm} b_{ij}\in {\Bbb C}$$
and $\sum _{j=1}^N b_{ij}b_{kj}^*=\delta_{ik}$ (with $\delta_{ik}$
the usual Kronecker's symbol ). We call a quantity an invariant
associated with the state $|\Psi\ra$ if it is invariant under all
local unitary transformations of $U_1\otimes U_2\otimes U_3$. By
generalizing the results of analysis on invariants for qubits
\cite{Linden}, the following quantities are invariants \cite{5}
under local unitary transformations:
$$
\begin{array}{ll}
I_0=\displaystyle\sum _{i,j,k=1}^N a_{ijk}a_{ijk}^*,&
I_1=\displaystyle\sum _{i,j,k,p,q,m=1}^N a_{ijk}a_{ijm}^*
a_{pqm}a_{pqk}^*,\\[3mm]
I_2=\displaystyle\sum _{i,j,k,p,q,m=1}^N a_{ikj}a_{imj}^*
a_{pmq}a_{pkq}^*,\hspace{0.2cm}& I_3=\displaystyle\sum
_{i,j,k,p,q,m=1}^N a_{kij}a_{mij}^* a_{mpq}a_{kpq}^*.
\end{array}
$$
A generalized concurrence is defined by \cite{5},
\begin{equation}
\label{2}
\begin{array}{crl}
C_N^3&=&\displaystyle\sqrt{\frac{N}{3(N-1)}(3I_0^2-I_1-I_2-I_3)}\\
&=&\displaystyle\sqrt{\frac{N}{6(N-1)}\sum (|a_{ijk}a_{pqm}-
a_{ijm}a_{pqk}|^2
+|a_{ijk}a_{pqm}-a_{iqk}a_{pjm}|^2+|a_{ijk}a_{pqm}-a_{pjk}a_{iqm}|^2)}.
\end{array}
\end{equation}
First, we prove the following result:

{\bf Lemma 1.} $C_N^3=0$ if and only if $|\Psi\ra$ is separable.

{\bf Proof.} It is clear that $C_N^3=0$ when $|\Psi\ra$ is
factorizable, i.e., when $$ a_{ijk}=a_i b_j c_k, \hspace{0.2cm}
{\rm  for\ some }\  a_i, b_j, c_k \in {\Bbb C}.$$ Conversely,
because $|\Psi\ra\neq 0$, there exists  $p_0, q_0, m_0$  such that
$a_{p_0q_0m_0}\neq 0$.  Hence from
$|a_{ijk}a_{pqm}-a_{ijm}a_{pqk}|=0$ we have $a_{ijk}=a_{ij} b_k$,
for some $a_{ij}, b_k \in {\Bbb C}$. Further we get $a_{ijk }=a_i'
b_j' c_k', \hspace{0.2cm} a_i',  b_j',  c_k' \in {\Bbb C}$. $\Box$

Let $\rho$ be a rank two state on $H\otimes H\otimes H$, with
$|E_1\ra, |E_2\ra$ being its two orthonormal
 eigenvectors corresponding to the two nonzero eigenvalues:
\begin{equation}
\label{3} \rho=p|E_1\ra\la E_1|+q|E_2\ra\la E_2|,
\end{equation}
where $q=1-p\in (0, 1)$. Generally $$ |E_{s_1}\ra=\sum_{i,j,k=1}^N
a_{ijk}^{s_1} e_i\otimes e_j\otimes e_k, \hspace{0.2cm} a_{ijk}^{s_1}\in
{\Bbb C},$$
with normalization $\sum_{i,j,k=1}^N a_{ijk}^{s_1}
(a_{ijk}^{s_1})^*=1, s_1=1, 2$.

Using Lemma 1, we have that $|\Psi\ra=\sum_{i,j,k=1}^N
a_{ijk}e_i\otimes e_j\otimes e_k$  is separable if and only if
$C_N^3=0$, i.e.,
\begin{equation}
\label{4} a_{ijk}a_{pqm}= a_{ijm}a_{pqk}, \hspace{0.2cm}
a_{ijk}a_{pqm}= a_{iqk}a_{pjm}, \hspace{0.2cm} a_{ijk}a_{pqm}=
a_{pjk}a_{iqm}, \hspace{0.2cm} \forall i, j, k, p, q, m.
\end{equation}
We adopt the notation
\begin{equation}
\label{5}
\begin{array}{l}
\alpha_1^{I}=a_{ijk}^2 a_{pqm}^2-a_{ijm}^2 a_{pqk}^2,
\hspace{0.2cm} \alpha_2^{I}=a_{ijk}^2 a_{pqm}^2-a_{iqk}^2
a_{pjm}^2, \hspace{0.2cm} \alpha_3^{I}=a_{ijk}^2
a_{pqm}^2-a_{pjk}^2 a_{iqm}^2, \\[3mm]
\beta_1^{I}=a_{ijk}^2 a_{pqm}^1+a_{ijk}^1
a_{pqm}^2-a_{ijm}^2 a_{pqk}^1-a_{ijm}^1 a_{pqk}^2,
\hspace{0.2cm}\\[3mm] \beta_2^{I}=a_{ijk}^2
a_{pqm}^1+a_{ijk}^1
a_{pqm}^2-a_{iqk}^2 a_{pjm}^1-a_{iqk}^1 a_{pjm}^2,\\[3mm]
\beta_3^{I}=a_{ijk}^2 a_{pqm}^1+a_{ijk}^1
a_{pqm}^2-a_{pjk}^2 a_{iqm}^1-a_{pjk}^1 a_{iqm}^2,\\[3mm]
\gamma_1^{I}=a_{ijk}^1 a_{pqm}^1-a_{ijm}^1
a_{pqk}^1,\hspace{0.2cm} \gamma_2^{I}=a_{ijk}^1
a_{pqm}^1-a_{iqk}^1 a_{pjm}^1, \hspace{0.2cm}
\gamma_3^{I}=a_{ijk}^1 a_{pqm}^1-a_{pjk}^1 a_{iqm}^1.
\end{array}
\end{equation}
where $I=\{i, j, k, p, q, m\},\ \forall i,j,k,p,q,m\in \{1,2,\dots,N\}$.

A vector of the form $|E_1\ra+\lambda |E_2\ra, \lambda \in {\Bbb
C}$,  is separable if and only if $\lambda$ is a common
 root of the following equation set $Eq_s^I:$
\begin{equation}
\label{6}
\begin{array}{l}
\alpha_s^I \lambda^2+\beta_s^I\lambda+\gamma_s^I=0.
\end{array}
\end{equation}
where $s=1,2,3$, and $I=\{i,j,k,p,q,m\}, \ \forall i,j,k,p,q,m\in
\{1,2,\dots,N\}$.

{\bf Lemma 2.} If $|E_2\ra$ is not separable, then $\rho$ is
separable if and only if (6) have two distinct roots.

{\bf Proof.} Suppose that $\rho=\sum_{t=1}^l p_t' |U_t\ra\la
U_t|$, with $l$ some positive integer and  $0< p_t'<1, \ \sum
p_t'=1,\ |U_t\ra$ being separable, $\forall t$. We can write them
as linear combinations of the two eigenvectors$|E_1\ra$
and$|E_2\ra$ which span the range of $\rho:\  |U_t\ra=c_1^t
|E_1\ra+c_2^t |E_2\ra$ (for some $c_1^t,c_2^t \in {\Bbb C}$). As
$|U_t\ra\neq 0, c_1^t, c_2^t$ can not be all 0. Without losing
generality, let $c_1^t\neq 0$. $|U_t\ra$  is then of the form
$|E_1\ra+\lambda_t |E_2\ra, \lambda_t=c_2^t/c_1^t$. From Lemma 1
$|U_t\ra$ is separable if and only if the parameter $\lambda_t$ is
a common root of the corresponding equation set (\ref{6}),
$\alpha_s^I \lambda_t^2+\beta_s^I\lambda_t+\gamma_s^I=0$.

Because $|E_2\ra$ is not separable, not all $\lambda_t's$ can be
equal, otherwise all the $U_t's$ would be constant multiples of a
fixed vector, and $\rho$ would be rank 1. On the other hand, as
$|E_2\ra$ is not separable, then $C_N^3$ is not zero. Hence there
is some $I_0, s_0$ such that $\alpha_{s_0}^{I_0}\neq 0$, i.e., the
relation $Eq_s^I$ is indeed a quadratic equation. It must have
exactly two roots, and so there are two values that are the only
possible choices for the $\lambda_t's$. But in order that there is
not only one possible choice, the above two roots must be
different.  And all the relations $Eq_s^I$ have these two
different roots.

Let $\mu_1,\mu_2$ be two distinct roots, which are common to all
of the equations $Eq_s^I$. Each vector  $|U_t\ra$ is either of the
form $|E_1'\ra=(|E_1\ra+\mu_1 |E_2\ra)/\sqrt{1+|\mu_1|^2}$, or of
the form $|E_2'\ra=(|E_1\ra+\mu_2 |E_2\ra)/\sqrt{1+|\mu_2|^2}$.

Therefore we can write $\rho$ as $\rho=p'|E_1'\ra\la
E_1'|+(1-p')|E_2'\ra\la E_2'|$ with $0< p'<1$. Comparing the
coefficients of $|E_k\ra\la E_l|, k, l=1,2$, with the ones in the
expression (3), we get that the following two  relations:
\begin{equation}
\label{7} \frac{p'}{1+|\mu_1|^2}+\frac{1-p'}{1+|\mu_2|^2}=p
\end{equation}
\begin{equation}
\label{8} \frac{\mu_1p'}{1+|\mu_1|^2}+\frac{\mu_2
(1-p')}{1+|\mu_2|^2}=0
\end{equation}
Solving for $p$ and $p'$ we get
\begin{equation}
\label{9}
p=(1-\mu_1\mu_2\frac{\bar{z}}{z})^{-1}, \hspace{0.2cm}
p'=\frac{\mu_2(1+|\mu_1|^2)}{z-\mu_1\mu_2\bar{z}}, \hspace{0.2cm}
{\rm where}\ z=\mu_2-\mu_1.
\end{equation}
Conversely, let $\mu_1, \mu_2$ be two distinct roots, which are
common to all of the equations $Eq_s^I$. From above discussion we
have $\rho=p'|E_1'\ra\la E_1'|+(1-p')|E_2'\ra\la E_2'|$, i.e.,
$\rho$ is separable. $\Box$

We first deal with the case where $a_{ijk}^{s_1}$ are all real. The
following conclusion is easily verified.

{\bf Lemma 3.} For a quadratic equation $a x^2+bx+c=0$ with
$a,b,c\in {\Bbb R}, a\neq 0$, and roots $\alpha, \beta$ with
$\alpha\neq \beta$,
$\gamma=(\bar{\alpha}-\bar{\beta})/(\alpha-\beta)$  is either $1$
or $-1$.

{\bf Theorem 1.} If all $a_{ijk}^{s_1}$ are real, $\rho$ is
separable if and only if one of the following
quantities($\triangle_1$ or $\triangle_2$) is zero:
\begin{equation}
\label{10}
\triangle_1=\sum
|\gamma_s^I-(1-p^{-1})\alpha_s^I|^2+\sum
|\beta_s^I\alpha_{s'}^{I'}-\alpha_s^I\beta_{s'}^{I'}|^2,
\end{equation}
\begin{equation}
\label{11}
\triangle_2=\sum
|\gamma_s^I+(1-p^{-1})\alpha_s^I|^2+\sum |\beta_s^I|^2,
\end{equation}
or, equivalently, one of the following two sets of relations (12) or (13) hold:
\begin{equation}
\label{12}
\gamma_s^I=(1-p^{-1})\alpha_s^I,
\hspace{0.2cm}
 \beta_s^I\alpha_{s'}^{I'}=\alpha_s^I\beta_{s'}^{I'}
\end{equation}
\begin{equation}
\label{13}
\gamma_s^I=-(1-p^{-1})\alpha_s^I,
\hspace{0.2cm}
 \beta_s^I=0
\end{equation}
where $s, s'=1, 2, 3$,  and $I, I'=\{i,j,k,p,q,m\},\ \forall
i,j,k,p,q,m\in \{1,2,\dots, N\}$.

{\bf Proof.} We prove the necessity part of the theorem in two
cases:

Case 1. $a).$  $|E_2\ra$ is not separable. We get that (6) have
two distinct roots from Lemma 2. These two roots are the solutions
to all the relations $Eq_s^I$. Consider for any  $ s=1,2,3,\
I=\{i, j, k, p, q, m\}, \ \forall i,j,k,p,q,m\in \{1,2,\dots,
N\}$,

1) if $\alpha_s^I\neq 0$, the corresponding relation (6) is not an
identity. All the quadratic equations in the set $Eq_s^I$ have the
same two distinct roots. From the standard  theory  of  quadratic
equations, we have
\begin{equation}
\label{12p}
\begin{array}{l}
\beta_s^I\alpha_{s_0}^{I_0}=\beta_{s_0}^{I_0}\alpha_s^I,
\end{array}
\end{equation}
\begin{equation}
\label{13p}
\begin{array}{l}
\gamma_s^I\alpha_{s_0}^{I_0}=\gamma_{s_0}^{I_0}\alpha_s^I.
\end{array}
\end{equation}

2) if $\alpha_s^I =0$, then the equations $Eq_s^I$ become
identities, i.e.,  $\beta_s^I$ and $\gamma_s^I$ must be $0$ too,
because otherwise at least one of the relations $Eq_s^I$ would be
a linear equation, and there would be no two distinct roots. Thus
in this case (12), (13) also hold.

$b).$  Because all $a_{ijk}^{s_1}$ are real number,  $\mu_1$ and
$\mu_2$ are roots of a quadratic equation with real coefficients.
From Lemma 3, $\mu_1\mu_2 =1-p^{-1}$ or $-(1-p^{-1})$. Since
$\mu_1\mu_2$ is real, the solution for $p'$ in (9) implies that
$\mu_2/(\mu_2-\mu_1)$ is real, which is possible if and only if
either the roots are both real or the roots are both purely
imaginary. In the first case, let $\mu_2> \mu_1$, we have
$\mu_1\mu_2=1-p^{-1}$. From (9), we get the condition that
$p'\in[0,1]$, which is equivalent to $\mu_2>0, \mu_1<0$. In  the
second case, we have $\mu_1\mu_2=-(1-p^{-1})$. The condition for
having purely imaginary roots of quadratic equations gives that
$\beta_s^I=0, \forall I, \ \forall s$.

$c).$ Finally, we observe that $\mu_1\mu_2$ is nothing but the
ratio $\gamma_{s_0}^{I_0}/ \alpha_{s_0}^{I_0}$, which is either $1-p^{-1}$
or $-(1-p^{-1})$. Therefore  we conclude that either
$\gamma_s^I=(1-p^{-1})\alpha_s^I$ or
$\gamma_s^I=-(1-p^{-1})\alpha_s^I$ for any $I$ and $s=1,2,3$.
Relation (10) is verified.

Case 2. $|E_2\ra$ is separable. In this case from (4), we have
$\alpha_s^I=0, \forall I, \forall s$. Since not all of the  $|U_t\ra$ can
be multiples of $|E_2\ra$, we must have at least one choice of
$\lambda$ such that $|E_1\ra+\lambda |E_2\ra$ is separable. This
must be a common root to all equations $Eq_s^I$ as before. All
these equations are now linear ones. When all
$\beta_s^I=\gamma_s^I=0$, it is easy to see that $|E_1\ra$ is
separable. Excluding this case, we see that there is only one
 possible choice of $\lambda$. Then $\rho$ can be expressed as
$$\rho=p''|E_2\ra\la E_2|+(1-p'')\frac{|E_1+\lambda E_2\ra\la
E_1+\lambda E_2|}{1+|\lambda|^2}.$$ That is $p''=1$, which is a
contradiction. Thus, if $|E_2\ra$ is separable, $|E_1\ra$ must be
separable too. It is clear that in this case (10) and (11) hold.

Now we prove the sufficiency part for the theorem. If (10) or (11)
holds, it is clear the equations $Eq_s^I$ have common roots. If
$|E_2\ra$ is not separable, then not all of these equations are
identities. And there are at most two common roots. If (10) holds,
the product of the two roots must be $1-p^{-1}<0$, so that the two
roots are real and unequal. If (11) holds, the two roots must be
purely imaginary. So in these two cases, we get that $\rho$ is
separable in terms of Lemma 2. If $|E_2\ra$ is separable, from
(10) or (11) we know $|E_1\ra$ is separable too and $\rho$ is separable. $\Box$

Generalizing the results in Theorem 1, we have, for the complex
$a_{ijk}^{s_1}$,

{\bf Theorem 2.} $\rho$ is separable if and only if there is
$\theta\in {\Bbb R}$ such that
\begin{equation}
\label{14}
\begin{array}{l}
\gamma_s^I=e^{i\theta}(1-p^{-1})\alpha_s^I, \hspace{0.2cm}
\beta_s^I\alpha_{s'}^{I'}=\alpha_s^I\beta_{s'}^{I'}
\end{array}
\end{equation}
where $s, s'=1,2,3$, and $I, I'=\{i,j,k,p,q,m\}, \ \forall
i,j,k,p,q,m\in \{1,2, \dots, N\}$,
 and
\begin{equation}
\label{15} \frac{\mu_2(1+|\mu_1|^2)}{z-\mu_1\mu_2\bar{z}}\in
[0,1],
\end{equation}
where $z=e^{i\theta}\bar{z}, z=\mu_2-\mu_1\neq 0$, $\mu_1$ and
$\mu_2$ are the roots of the equation
$\alpha_s^I\lambda^2+\beta_s^I\lambda+\gamma_s^I=0$ for some $I$
and $s$ such that $\alpha_s^I\neq 0$.

{\bf  Proof.} The proof of necessity is similar to the proof of
the corresponding part in Theorem 1. One only needs to note that
since $z/\bar{z}$ is of modulus 1, a phase factor $e^{i\theta}$
appears in this case.

Now if (14) holds, it is clear that the equations $Eq_s^I$ have
common roots. If $|E_2\ra$ is not separable, then  some of the
$\alpha_s^I$ are nonzero. The corresponding equations $Eq_s^I$
have exactly two roots which are different by condition (15).
Therefore $\rho$ is separable from Lemma 2. If $|E_2\ra$ is
separable, by (14) we know that all $\gamma_s^I$ are $0$. Hence both
$|E_2\ra$ and $|E_1\ra$ are separable, and so is $\rho$. $\Box$

{\bf Corollary.} Let $|E_2\ra$ be the maximally entangled vector
given by $|E_2\ra=(1/\sqrt{N})\sum_{i=1}^N e_i\otimes e_i\otimes
e_i$. For any vector $|E_1\ra$ which is orthogonal to $|E_2\ra$,
$\rho=p|E_1\ra \la E_1|+(1-p) |E_2\ra \la E_2|$ is not separable
for $0<p<1/2$.

{\bf Proof.} Let $$
C_{(1)}=\sqrt{\frac{N}{6(N-1)}\sum_{I,s}|\gamma_s^I|^2}
\hspace{0.2cm}~ {\rm and }~ \hspace{0.2cm}
C_{(2)}=\sqrt{\frac{N}{6(N-1)}\sum_{I,s}|\alpha_s^I|^2}$$ be the
generalized concurrences associated with the states $|E_1\ra$ and
$|E_2\ra$, respectively, where $s=1,2,3, \ I=\{i,j,k,p,q,m\}, \
\forall i,j,k,p,q,m\in \{1,2, \dots, N\}$.

Suppose $\rho$ is separable. From the necessary condition for
separability,  $\gamma_s^I=e^{i\theta}(1-p^{-1})\alpha_s^I$, we
get $C_{(1)}=\frac{1-p}{p}C_{(2)}$. As $|E_2\ra$ is maximally
entangled, $C_{(2)}\neq 0$, and $C_{(1)}/C_{(2)}=\frac{1-p}{p}\leq
1$, so we have $p\geq 1/2$, which is a contradiction. $\Box$

The above approach can be extended to the case of multiquantum
systems. We consider now the separability of $|\Psi_M\ra$ on $M \
\ N$-dimensional quantum systems, where
\begin{equation}
\label{16}
|\Psi_M\ra=\sum_{i_1,i_2,\dots, i_M=1}^{N} a_{i_1 i_2
\dots i_M}e_{i_1}\otimes e_{i_2}\otimes\dots\otimes e_{i_M},
\hspace{0.2cm}a_{i_1 i_2\dots i_M}\in {\Bbb C}
\end{equation}
with $\sum a_{i_1 i_2\dots i_M}a_{i_1 i_2 \dots i_M}^*=1$. We have
a quadratic $I_0=\sum a_{i_1 i_2 \dots i_M}a_{i_1 i_2 \dots
i_M}^*$ and $d=2^{M-1}-1$ biquadratic invariants:
$$I_{TS}=\sum a_{TS}a_{TS'}^*a_{T'S'}a_{T'S}^* $$
where $T$ and $T'$ are all possible  nontrivial subset of
$I=\{i_1,i_2,\dots, i_M \}, I'=\{\tilde{i_1}, \tilde{i_2}, \dots,
\tilde{i_M}\}$, respectively, $\forall i _k,
\tilde{i_k}=1,2,\cdots, N$, $k=1, 2, \dots, M$ (i.e., $T\neq \phi$
and $T\neq I$), $S=I\backslash T, S'=I'\backslash T'$.

$T$ and $T'$ are subindices of $a$, associated with the same
position. A generalized concurrence can be defined by
\begin{equation}
\label{17}
C_N^M=\sqrt{\frac{N}{d(N-1)}(dI_0^2-I_1-I_2-\cdots-I_d)}=\sqrt{\frac{N}{4d(N-1)}\sum_p
|a_{TS}a_{T'S'} -a_{TS'}a_{T'S}|^2}
\end{equation}
where $\sum_p$ stands for the summation over all possible
combination of the indices of $T$ and $S$. Similar to Lemma 1, one
can prove:

{\bf Lemma 4.} $C_N^M=0$ if and only if $|\Psi_M\ra$ is separable.

Let $\rho$ be a rank two state on $H\otimes H\otimes\cdots\otimes
H$, with $|E_1\ra, |E_2\ra$ being its two orthonormal eigenvectors
corresponding to the two nonzero eigenvalues:
\begin{equation}
\label{18}
\begin{array}{l}
\rho=p|E_1\ra\la E_1|+q|E_2\ra\la E_2|,
\end{array}
\end{equation}
where $q=1-p\in (0,1)$. Generally $$ |E_{s_1}\ra=\sum_{i_1 i_2\dots
i_M=1}^N a_{i_1 i_2 \dots i_M}^{s_1} e_{i_1}\otimes
e_{i_2}\otimes\dots\otimes e_{i_M}, \hspace{0.2cm}a_{i_1i_2\dots
i_M}^{s_1} \in {\Bbb C}, $$ with normalization  $\sum a_{i_1 i_2 \dots
i_M}^{s_1} (a_{i_1 i_2 \dots i_M}^{s_1})^*=1, s_1=1,2.$

Using Lemma 4, we have  $|\Psi_M\ra$ is separable if and only if
\begin{equation}
\label{19}
\begin{array}{l}
a_{TS}a_{T'S'}=a_{TS'}a_{T'S},
\end{array}
\end{equation}
where $T$ and $T'$ are all possible  nontrivial subset of
$I=\{i_1,i_2,\dots, i_M \}, I'=\{\tilde{i_1}, \tilde{i_2}, \dots,
\tilde{i_M}\}$, respectively, $\forall i _k,
\tilde{i_k}=1,2,\dots,N$, $k=1, 2, \dots, M$ (i.e., $T\neq \phi$ and $T\neq I$),
$S=I\backslash T, S'=I'\backslash T'$.

With the notations:
$$
\begin{array}{l}
\alpha_{TS}^{T'S'}=a_{TS}^2 a_{T'S'}^2-a_{TS'}^2 a_{T'S}^2,
\hspace{0.2cm}
\gamma_{TS}^{T'S'}=a_{TS}^1 a_{T'S'}^1-a_{TS'}^1 a_{T'S}^1,\\[3mm]
\beta_{TS}^{T'S'}=a_{TS}^2 a_{T'S'}^1+a_{TS}^1
a_{T'S'}^2-a_{TS'}^2 a_{T'S}^1-a_{TS'}^1 a_{T'S}^2,
\end{array}
$$ we have that a vector of the form $|E_1\ra+\lambda |E_2\ra,
\lambda \in {\Bbb C}$, is separable if and only if   $\lambda$ is
a common root of the following equation set:
\begin{equation}
\label{20}
\begin{array}{l}
Eq_{TS}^{T'S'}:\
\alpha_{TS}^{T'S'}\lambda^2+\beta_{TS}^{T'S'}\lambda+\gamma_{TS}^{T'S'}=0.
\end{array}
\end{equation}
Similar to the case $M=3$, one has:

{\bf Lemma 5.} $\rho$ is separable if and only if (20) have two
distinct roots.

From Lemma 4 and Lemma 5 it is straightforward to prove the
following conclusion:

{\bf Theorem 3.} If all $a_{i_1i_2\dots i_M}^{s_1}$ are real,
$\rho$ is separable if and only if one of the following quantities
($\triangle_1$ or $\triangle_2$) is zero:
\begin{equation}
\label{21} \triangle_1=\sum
|\gamma_{TS}^{T'S'}-(1-p^{-1})\alpha_{TS}^{T'S'}|^2+ \sum
|\beta_{TS}^{T'S'}\alpha_{T_1S_1}^{T_1'S_1'}-\alpha_{TS}^{T'S'}\beta_{T_1S_1}^{T_1'S_1'}|^2,
\end{equation}
\begin{equation}
\label{22}
\triangle_2=\sum
|\gamma_{TS}^{T'S'}+(1-p^{-1})\alpha_{TS}^{T'S'}|^2+\sum
|\beta_{TS}^{T'S'}|^2,
\end{equation}
or, equivalently, one of the following two sets of relations (23) or (24) hold:
\begin{equation}
\label{23}
\gamma_{TS}^{T'S'}=(1-p^{-1})\alpha_{TS}^{T'S'},
\hspace{0.2cm}
 \beta_{TS}^{T'S'}\alpha_{T_1S_1}^{T_1'S_1'}=\alpha_{TS}^{T'S'}\beta_{T_1S_1}^{T_1'S_1'}
\end{equation}
\begin{equation}
\label{24}
\gamma_{TS}^{T'S'}=-(1-p^{-1})\alpha_{TS}^{T'S'},
\hspace{0.2cm}
 \beta_{TS}^{T'S'}=0
\end{equation}
where $T$ and $T'$ are all possible  nontrivial subset of
$I=\{i_1,i_2,\dots, i_M \}, I'=\{\tilde{i_1}, \tilde{i_2}, \dots,
\tilde{i_M}\}$, respectively, $\forall i _k, \tilde{i_k}=1,2,
\dots, N$, $k=1,2,\dots, M$ (i.e., $T\neq \phi$ and $T\neq I$), $S=I\backslash T,
S'=I'\backslash T'$, $T_1$ and $T_1'$ are all possible nontrivial
subset of $J=\{j_1,j_2,\dots, j_M \}, J'=\{\tilde{j_1},
\tilde{j_2}, \dots, \tilde{j_M}\}$, respectively, $\forall j _k,
\tilde{j_k}=1,2,\dots, N$, $k=1,2,\dots, M$  (i.e., $T_1\neq \phi$ and $T_1\neq J$),
$S_1=J\backslash T_1, S_1'=J_1'\backslash T_1'$.

Extending Theorem 3 to general complex coefficients
$a_{i_1i_2\dots i_M}^{s_1}$, we have

{\bf Theorem 4.} $\rho$ is separable if and only if there is
$\theta\in {\Bbb R}$ such that
 \begin{equation}
\label{25}
\gamma_{TS}^{T'S'}=e^{i\theta}(1-p^{-1})\alpha_{TS}^{T'S'},
\hspace{0.2cm}
 \beta_{TS}^{T'S'}\alpha_{T_1S_1}^{T_1'S_1'}=\alpha_{TS}^{T'S'}\beta_{T_1S_1}^{T_1'S_1'}
\end{equation}
\begin{equation}
\label{26} \frac{\mu_2(1+|\mu_1|^2)}{z-\mu_1\mu_2\bar{z}}\in
[0,1].
\end{equation}
where $T,T',S,S',T_1,T_1',S_1,S_1'$ are defined  as in Theorem 3,
$z=e^{i\theta}\bar{z}, z=\mu_2-\mu_1\neq 0$, $\mu_1$ and $\mu_2$
are the roots of the equation
$\alpha_{TS}^{T'S'}\lambda^2+\beta_{TS}^{T'S'}\lambda+\gamma_{TS}^{T'S'}=0$
for some $ T, S,T' , S'$ such that $\alpha_{TS}^{T'S'}\neq 0$.

For a given rank two density matrix  on $H\otimes
H\otimes\cdots\otimes H$,  to find its separability one only needs
to calculate the two eigenvectors $|E_1\ra,|E_2\ra$ corresponding
to the two nonzero eigenvalues and check if formula (25) is
satisfied or not.

{\bf Corollary.} Let $|E_2\ra$ be the maximally entangled vector
given by $|E_2\ra=(1/\sqrt{N})\sum_{i=1}^N e_i \otimes e_i\otimes
\cdots \otimes e_i$. For any vector $|E_1\ra$ which is orthogonal
to $|E_2\ra$, $\rho =p|E_1\ra\la E_1|+(1-p)|E_2\ra\la E_2|$ is not
separable for $0<p<1/2$.

We have studied the sufficient and necessary conditions for
separability of rank two mixed states in higher-dimensional
quantum systems on $H\otimes H\otimes\dots\otimes H$. The
separability condition for these kind of mixed states in arbitrary
dimensions is explicitly given. A nonseparability inequality is
also given for the case where one of the eigenvectors
corresponding to nonzero eigenvalues of a density matrix is
maximally entangled. The results can be generalized to the case of
rank two mixed states on $H_1\otimes H_2\otimes\dots\otimes H_M$,
where $H_i$, $i=1,...,M$, may have different dimensions.

\noindent{\bf Acknowledgement} The work is supported by the NSF of
China (No.19975061) and the National Key Project for Basic
Research of China (G1998030601).


\vskip 8mm
\begin{thebibliography}{20}

\bibitem{effects}
A. Ekert, Phys. Rev. Lett. {\bf 67}, 661 (1991).

\bibitem{geste}
C. H. Bennett  and S. J. Wiesner, Phys. Rev. Lett. {\bf 69}, 2881
(1992).

\bibitem{Tel}
C. Bennett, G. Brassard, C. Crepeau, R. Jozsa, A. Peres and W. K.
Wootters, Phys. Rev. Lett.  {\bf 70}, 1895  (1993).

\bibitem{peresbook} A. Peres, ``Quantum Theory: Concepts and Methods'',
Kluwer Academic Publishers (1995).

\bibitem{primer}
see M. Horodecki, P. Horodecki and R. Horodecki in ``Quantum
Information - Basic Concepts and Experiments'', Eds. G. Alber and
M. Weiner, (Springer, Berlin, 2000).\\
M. Lewenstein, D. Bru{\ss}, J. I. Cirac, B. Kraus, M. Ku\'s, J.
Samsonowicz, A. Sanpera,  and R. Tarrach, J. Mod. Phys. {\bf 47},
2481 (2000).

\bibitem{Werner} R. Werner, Phys. Rev. A{\bf 40}, 4277 (1989).

\bibitem{Peres} A. Peres  Phys. Rev. Lett. {\bf 77}, 1413 (1996).

\bibitem{ho96} M. Horodecki, P. Horodecki, and R. Horodecki,
Phys. Lett. A {\bf 223}, 8 (1996).

\bibitem{tran} P. Horodecki Phys. Lett. A {\bf 232}, 333 (1997).

\bibitem{rew2000} D. Bouwmeester, A. Ekert, and
A. Zeilinger (Eds.), ``The Physics of Quantum Information",
(Springer, Heidelberg, 2000), chapter 6 and references therein.

\bibitem{four}
W. D\"ur, G. Vidal and J.I. Cirac, Phys. Rev. A {\bf 62},
062314(2000).\\
V. Coffman, J. Kundu ans W. Wootters, Phys. Rev. A
{\bf 61}, 052306(2000).\\
W. Wootters, Phys. Rev. A {\bf 63}, 052302(2001).

\bibitem{sharing} W. Tittel, H. Zbinden and N. Gisin,
{\it Quantum secret sharing using pseudo-GHZ states},
quant-ph/9912035.

\bibitem{exper}
A. Zeilinger, M. A. Horne, H. Weinfurter, and M. \.Zukowski, Phys.
Rev. Lett. {\bf 78}, 3031 (1997).\\
D. Bouwmeester, J-W. Pan, M. Daniell, H. Weinfurter, and A.
Zeilinger, Phys. Rev. Lett. {\bf 82}, 1345 (1999).\\
A. S\o rensen  and K. M\o lmer, Phys. Rev. Lett. {\bf 82},
1971(1999).\\
C. A. Sackett, D. Kielpinski, B. E. King, C. Langer, V. Meyer, C.
J. Myatt, M. Rowe, Q. A. Turchette, W. M. Itano, D. J. Wineland,
and C. Monroe, Nature {\bf 404}, 256 (2000).\\
D. Kielpinski, A. Ben--Kish, J. Britton, V. Meyer, M.A. Rowe, C.
A. Sackett, W. M. Itano, C. Monroe, and D. J. Wineland, {\it
Recent Results in Trapped-Ion Quantum Computing},
quant-ph/0102086.

\bibitem{persis}
H.J. Briegel and R. Raussendorf, Phys. Rev. Lett. {\bf 86},
910(2001).

\bibitem{schmidt}
J. Eisert and H.J. Briegel, Phys. Rev. A {\bf 64}, 022306(2001).

\bibitem{22n}
S. Karnas and M. Lewenstein, {\it Separability and entanglement in
${\cal C}^2\otimes{\cal C}^2\otimes{\cal C}^N$  composite quantum
systems}, quant-ph/0102115.

\bibitem{4} Albeverio S., Fei S. M. and Goswami, D., Separability of
rank two quantum states, Phys. Lett. A 286(2001)91-96.\\
see also P. Horodecki, J.A. Smolin, B.M. Terhal and A.V.
Thapliyal, {\it Rank Two Bipartite Bound Entangled States Do Not
Exist}, quant-ph/9910122.\\
P. Horodecki, M. Lewenstein, G. Vidal and I. Cirac, Phys. Rev. A
{\bf 62}, 032310 (2000).

\bibitem{Linden} N. Linden and P. Popescu, Fortsch.
Phys. {\bf 46}, 567 (1998).

\bibitem{5} Albeverio S., Fei S. M.,  J. Opt. B: Quantum Semiclass. Opt. 3,
223(2001).

\end{thebibliography}
\end{document}